\documentstyle[preprint,aps]{revtex}

\newcommand{\beq}{\begin{equation}}
\newcommand{\eeq}{\end{equation}}
\newcommand{\beqa}{\begin{eqnarray}}
\newcommand{\eeqa}{\end{eqnarray}}
\begin{document}

\draft

\title{Quantum simulations of the superfluid-insulator transition\\
for two-dimensional, disordered, hard-core bosons}

\author{Shiwei Zhang and N.~Kawashima}
\address{Center for Nonlinear Studies and Theoretical Division\\
Los Alamos National Laboratory, Los Almaos, NM 87545}
\author{J.~Carlson and J.E.~Gubernatis}
\address{Theoretical Division,
Los Alamos National Laboratory, Los Almaos, NM 87545}

\date{\today}
\maketitle

\begin{abstract}
We introduce two novel quantum Monte Carlo methods and employ
them to study the superfluid-insulator transition in a two-dimensional
system of hard-core bosons.  One of the methods is appropriate
for zero temperature and is based upon Green's function Monte Carlo; the other
is a finite-temperature world-line cluster algorithm.
In each case we find that the dynamical exponent is consistent with the
theoretical prediction of $z=2$ by Fisher and co-workers.
\end{abstract}



Recently, the study of the superfluid-insulator phase transition in
systems of disordered interacting bosons has attracted considerable
attention.  As discussed by Fisher and co-workers\cite{fisher89}, this
transition is associated with a zero-temperature transition from a
superfluid to an insulating state as the strength of the disorder is
increased. In addition to experimental studies in systems such as
$^4$He adsorbed in porous media, random magnets, Josephson junction
arrays, exciton lifetimes in quantum well structures, vortices in bulk
type-II superconductors, a huge amount of numerical work has been
performed on models of these systems \cite{zimanyi94}.

Computer simulations are a natural candidate for studying this
transition.  The results of the simulations to date, however, are
somewhat surprising as they seem to indicate that slightly different
models lead to different values of the relevant critical exponents.
The most significant disagreement exists between the numerical work on
the Vallain form of the disordered XY model
\cite{sorensen92,wallin94}, which has produced a value of the
dynamical exponent z consistent with the original prediction of $z=2$
given by Fisher and coworkers, and work on the hard-core boson model
\cite{makivic93}, which obtained $z=0.5$.  This difference was
suggested in the latter work to be the result of (unspecified)
low-lying excitations that are not encompassed in the models studied
by others.  Recently, these claims have been questioned
\cite{zimanyi94,wallin94,weichman94b}.  Most of the numerical work on
these dirty-boson systems has been based upon finite-temperature
quantum Monte Carlo (MC) methods, although exact diagonalization
\cite{runge92} and large-N methods\cite{weichman94a} have also been
employed.

The aim of this Letter is to introduce novel zero- and finite-temperature
quantum MC methods and to use them
to study the superfluid-insulator transition in a hard-core
dirty-boson system.  As we discuss below, these two methods have
important, distinct conceptual and practical advantages over the
standard algorithms.  Our zero-temperature algorithm is based on the
Green's function Monte Carlo (GFMC) method \cite{ceperley79} and allows one
to compute ground-state properties directly and to
perform scaling analysis in the spatial dimensions only.
The finite-temperature algorithm
employed is a new way to implement the world-line algorithm that can
reduce the equilibration and auto-correlation times by several orders
of magnitude.

Our Hamiltonian is the standard hard-core boson model with
random disorder:
\begin{equation}
  H = -J\sum_{\langle ij \rangle} (b_i^\dagger b_j + b_j^\dagger b_i)
    + \sum_i V_i n_i^2
\end{equation}
where $J$ is the overlap integral between neighboring bosons, $\langle
ij \rangle$ denotes a nearest neighbor pair of lattice sites,
$b_i^\dagger$ and $b_i$ are the boson creation and destruction
operators at lattice site $i$, and $n_i = b_i^\dagger b_i$ is the
boson number operator which is restricted to have the eigenvalues of 0
or 1. $V_i$ is the on-site disorder and is assumed to be randomly and
uniformly distributed over the interval $(-V,V)$.  We consider
$N_b$ bosons on a
two-dimensional square lattice of $L\times L$ ($=N$) sites,
with periodic boundary conditions in both spatial dimensions.
In this work we restrict ourselves to half-filling ($\rho=N_b/N=1/2$).

Two-dimensions occupies a special place in the theory of this model
because of the predicted existence of a finite universal value of the
dc conductivity \cite{fisher89}.  Scaling theory also predicts that
near the critical value of the disorder $V_c$ the singular part of the
total (normal plus superfluid) compressibility $\kappa$ is expected to
behave as $|\delta|^{\nu(z-d)}$, where $\delta = V-V_c$.  If $z=d=2$,
then near $V_c$ this quantity is finite and
independent of system size.  We
observed this behavior in our zero-temperature simulations.

To apply the GFMC method \cite{carlson89}, we define a kernel $K\equiv
c-H$. In an occupation number basis $R$, all matrix elements
$K(R,R^\prime)$ are readily obtained, and the constant $c$ is chosen
such that these elements are non-negative. GFMC projects out the
ground-state wavefunction $\psi_0$ by iterating the equation
$\psi^{(t+1)}=K\psi^{(t)}$ from any positive initial function
$\psi^{(0)} = \psi_T$. We can rewrite the equation in a form more
appropriate for the MC process:
\begin{equation}
 \tilde \psi^{(t+1)}(R)
 \propto\sum_{R^\prime} \tilde K(R,R^\prime) \tilde \psi^{(t)}(R^\prime),
\end{equation}
where $\tilde \psi^{(t)}(R)=\psi_T(R)\psi^{(t)}(R)$ and $\tilde
K(R,R^\prime)=\psi_T(R) K(R,R^\prime) \psi_T^{-1}(R^\prime)$.  The
trial wave function $\psi_T$ for the ground state serves as an
importance function. In the MC process, $\tilde \psi^{(t)}$ at each
$t$ is sampled by a finite ensemble of configurations (walkers) each
carrying a weight. A walker at $R^\prime$ at time $t$
advances to time $t+1$ by sampling $R$ from
$\tilde K(R,R^\prime)/\sum_R\tilde K(R,R^\prime)$, and determining a new weight
from the product of the old weight and $\sum_R \tilde K(R,R^\prime)$.
Splitting and combining procedures are applied at
times to control the weights.
A good $\psi_T$ can greatly reduce the fluctuations in the weight factors.

Once the iteration has reached the asymptotic limit, ground-state
properties can be calculated from the distribution of configurations.
The projection of the ground state here is free of the Trotter
approximation or other systematic errors.
The ground-state energy is given by $E=\left<  H\psi_T /\psi_T\right>$.
The compressibility $\kappa=(N_b {d^2E / dN_b^2})^{-1}$ can be
computed directly from $E(N_b)$ by finite differences.

In GFMC, calculations of quantities other than the energy can usually
be performed only approximately because the result of the simulation
is a distribution of configurations proportional to $\psi_T \psi_0$.
However, the
superfluid density is a rather special property, being directly
associated with the response of the system to moving boundaries:  it
is simply proportional to the long imaginary-time
diffusion distance of the world-lines.  By analogy with the arguments in
\cite{pollack87}, a diffusion displacement {\bf D} can be defined:
\begin{equation}
 {\bf D}(\tau)={1 \over L}
 \Bigl[\sum_{i=1}^{N_b}
  ({\bf r}_i^{(t+\tau)} - {\bf r}_i^{(t)})\Bigr],
\label{eq:displacement}
\end{equation}
where boundary crossings in the $x$- and $y$-directions are included.
It can then be shown that the superfluid density $\rho_s$ is given by
\begin{equation}
 \rho_s={c-E \over 4J} \lim_{\tau \rightarrow \infty}
{<{\bf D}^2(\tau)> \over \tau}.
\end{equation}
With GFMC, we are essentially examining the diffusion over a finite
portion of the (infinite) zero-temperature paths. In the full path
integral, the density of configurations would be proportional to
the ground-state wave function $\psi_0$ at each end of this
segment.  We replace this condition at one end by an approximate one in which
the density is proportional to $\psi_T$ instead of
$\psi_0$.  However, we emphasize that in the limit ($\tau
\rightarrow \infty$) that determines the superfluid density, these
edge conditions have no effect, and the method is exact.
We expect this method of computing
the superfluid density to be quite valuable in many other
applications, including the continuum.

To extract the superfluid density, it is necessary to extrapolate
diffusion distance to infinite $\tau$.  Due
to the local motion of the paths, even non-superfluid states will propagate
a finite diffusion distance at large $\tau$.  In order to project out
this localized effect, we use the following to model the
asymptotic diffusion
\begin{equation}
 \left<{\bf D}^2 (\tau) \right> / \tau =
a' + (b'/\tau) [1 -{\rm exp}(- c'\tau)],
\end{equation}
where the constant $a'$ determines the superfluid density and $b'$ and
$c'$ are associated with ``local'' diffusion. The
effects of higher ( $\ge 2)$ orders in $1/\tau$ are small and
different functional forms with the same asymptotic expansion
produce identical answers.

Our $\psi_T$ is the product of a one-body term and a Jastrow factor:
\begin{equation}
 \psi_T(R) =\Phi_1 \enskip {\rm exp}[-\sum_{m,n=1}^{N_b}f(r_{mn})],
\end{equation}
where $\Phi_1={\rm exp}(-a\sum_i n_i V_i-b \sum_{<ij>}n_iV_j)$.
The Jastrow factor contains long-range two-body correlations of the form
 \begin{equation}
 f(r_{mn})=\cases {\alpha,            &if $r_{mn}=1$ \cr
                   \beta/r_{mn}^\eta, & otherwise\cr},
\end{equation}
where $r_{mn}$ denotes the distance between bosons $m$ and $n$.  The
variational parameters, $a$, $b$, $\alpha$, $\beta$, and $\eta$, are
optimized by minimizing the variance of the local energy.
Due to the discrete nature of the problem, importance sampling with
the long-range correlation in $\psi_T$ can be implemented in a very
efficient way.  We have verified
that our results, in particular the superfluid density, remain
unchanged except in statistical errors when these parameters deviate
from their optimal values.

The results of our zero-temperature simulations are summarized in
Figs.~\ref{fig.1} and \ref{fig.2}.  We calculated the energy,
compressibility, and superfluid density for system sizes of $4\times
4$ up to $12 \times 12$.  The disorder magnitude $V$ is the
controlling parameter of the phase transition, and for each $V$, about
$100$ realizations of the disorder are studied.  We verified that
our results for the $4 \times 4$ system are in excellent agreement
with the exact diagonalization studies \cite{runge92}.
Figure~\ref{fig.1} presents the superfluid density as a function of
disorder $V$ for various system sizes.  The transition is evident
by the changes in $\rho_s$ with system size for various $V$.

To determine the precise scaling behavior, we used the
following scaling relation for the superfluid density:
\begin{equation}
  \rho_s(\delta,L) = L^{-z} Y(L^{1/\nu}\delta)
\label{eq:scaling0}
\end{equation}
where $Y$ is the scaling function.  From this form, we see that when
scaled with the correct value of $z$, curves for various $L$ will have
a point of common intersection that marks the critical point $V_c$
($\delta=0$).  Results for $z = 0.5$, $2.0$, and $3.0$ are shown in
Fig.~\ref{fig.2}.  Clearly, $z=2$ is much favored over the other two
values. A statistical analysis yields $V_c=9.9 \pm 0.4$, and
$z=2.0\pm 0.4$.  As noted previously, if the $z=2$ scaling behavior is
correct, the compressibility $\kappa$ should be finite and continuous
in the region of the transition.  We observed no statistically
significant change in $\kappa$ in this region (see Fig.~\ref{fig.1}
inset), and found $\kappa \sim 0.10(1)$.  We also determined the
critical exponent $\nu$ to be $0.9 \pm 0.1$.  This result is
consistent with the lower bound of unity predicted by theory and with
the value obtained for the quantum rotor model.  It is not consistent
with the estimate of $2.2 \pm 0.2$ found in
\cite{makivic93}.

To obtain evidence complementary to the zero-temperature
calculation, we also performed finite-temperature MC
simulations and employed a finite-size scaling analysis with two lengths,
the lattice size $L$ and the inverse-temperature $\beta$.

To produce data, we extended a recently developed cluster algorithm
\cite{kawashima94} which is a form of world-line MC found to have
smaller equilibration and autocorrelation times than the conventional
implementation.  The algorithm is based on two simple observations: in
the absence of disorder, the hard-core boson model in two-dimensions
maps onto a $S=1/2$ XY model \cite{matsubara57} and the world-line
algorithm for the $S=1/2$ XY model can be mapped onto the loop
algorithm \cite{evertz93} for the 6-vertex model \cite{wiese94}.  When
used for world-lines, this method replaces the local movement of the
world-lines by global changes of loops: the world-lines decompose into
loops, each of which can be ``flipped'' with a probability of 1/2.  In
the presence of disorder the flipping probability for a loop is
\begin{equation}
  1/[1+\exp(\case{1}{2}\Delta \tau \sum_{i\, \in\, \hbox{loop}}(1-2n_i)V_i]
\end{equation}
where $\Delta \tau$ is the imaginary-time spacing characteristic of the
world-line method.  This flipping probability and the loop
construction \cite{kawashima94,wiese94} define our algorithm.  Both the cluster
and the conventional method suffer very long equilibration and
autocorrelation times in strong disorder and at low temperature.
These facts are the limiting factors of the finite-temperature
simulations.

Because the winding number of each world-line is conserved in the
conventional method, somewhat complicated procedures are necessary for
measuring the superfluid density from the winding number variance
\cite{makivic93}.  In the new algorithm, however, winding
number ${\bf W}$ is not conserved, so we can directly measure the superfluid
density from $\rho_s = \langle {\bf W}^2 \rangle/4J\beta$ where
${\bf W}$ is ${\bf D}(\tau)$ in (\ref{eq:displacement}) evaluated at
$\tau=\beta$.

We performed simulations for two sets of space-time aspect ratios: one
with $\beta/L^2 = 1/4$ and the other with $\beta/L^{1/2} = 1/2$,
corresponding to the two predictions $z=2$ and $z=1/2$.  We used an
open boundary condition in the temporal direction in the latter case
to make it possible for the winding numbers to take non-integral
values.  Allowing this possibility is useful when the winding number
fluctuations are much smaller than unity, as in the case of the
second set of simulations.  The number of MC steps used was
up to $2\times 10^5$ for the severest case.  The number of random
samples for the average is 32 to 128.  We carefully monitored the
equilibration times by calculating a Hamming distance similar to the
one adopted in \cite{wallin94} and found that as the system size or
the inverse temperature increases, the equilibration time increases
rapidly.

The finite-size
scaling form for the finite-temperature superfluid density is
\beq
  \rho_{\rm S}(\delta,L,\beta)
       = \beta^{-1}\Upsilon\bigl( \delta L^{1/\nu}, \beta/L^z \bigr),
\label{eq:finiteT_scl}
\eeq
with the scaling function $\Upsilon$.
The critical value $V_c$ should be the same as
that obtained in the zero-temperature calculations.
In Fig.~\ref{fg:FiniteTemperature}, we present results
corresponding to the two sets of simulations.
Based on (\ref{eq:finiteT_scl}), curves should display a common
crossing at $V=V_c$ if the chosen aspect ratio corresponds to the
correct $z$.

We find that, contrary to the observation in \cite{makivic93}, a
common crossing point exists at $z=2$.  In addition, the location of
the estimated crossing point in (a) is $V_c = 9.78 \pm 0.35$, in good
agreement with the zero-temperature calculation.  An apparent crossing
occurs in (b) for $z=1/2$, but the crossing points have expanded to
fill a larger region.  Such ``false'' crossings can be difficult to
diagnose given a limited range of system sizes and temperatures.
Here, however, we note that the crossing region in (b) is very
different from that found in the zero-temperature simulation, which
should not be the case if $z = 1/2$.  Finally, we remark that an
analysis of the results in (a) yields $\nu = 0.90 \pm 0.13$.

In conclusion, this combination of zero- and finite-temperature
algorithms have proven very powerful in studies of the
superfluid-insulator transition in disordered hard-core bosons.
The GFMC algorithm we have used produces {\it exact}
results for the energy, compressibility, and
superfluid density. As a zero-temperature method, it
allows us to perform scaling in the spatial variables only,
eliminating the need for any {\it a priori} assumptions.
Our results for the compressibility and superfluid density
convincingly demonstrate that $z=2$.  This conclusion is
strongly supported by the finite-temperature
simulations, which are completely consistent with the zero-temperature
calculations.  The combination of zero-
and finite-temperature simulations allow for a quite
sensitive determination of the correct scaling behavior.
Further details concerning our algorithms and our
simulations will be reported elsewhere.

We acknowledge helpful conversations with D.~Ceperley, M.P.A.~Fisher,
M.H.~Kalos, R.T.~Scaletter, N.~Trevedi, and P.~Weichman.
We thank the Institute
for Theoretical Physics (Santa Barbara) for its hospitality while part
of the work was performed.  Most of the simulations were performed on
the computers at the National Energy Research Supercomputer Center and
the Cornell Theory Center. The work of J.E.G.\ and S.Z.\ was
supported in part by the High Performance Computing and Communication
program of the Department of Energy.

\begin{figure}
\caption{The superfluid density $\rho_s$ for different lattice sizes
as a function of the strength of the disorder. The inset shows the
corresponding compressibility $\kappa$ in the vicinity of $V_c$.}
\label{fig.1}
\end{figure}

\begin{figure}
\caption{
Scaling plots assuming (a) $z=2$, (b) $z=1/2$, and (c) $z=3$. The
common crossing point is evident for $z=2$. For $z=1/2$, the
crossing points systematically shift to the right as $L$ is increased.
For $z=3$, the shift is to the left.}
\label{fig.2}
\end{figure}

\begin{figure}
\caption{
Scaling plots with fixed aspect ratio assuming (a) $z=2$ and (b)
$z=1/2$.}
\label{fg:FiniteTemperature}
\end{figure}


\begin{references}


\bibitem{fisher89}
M.P.A.~Fisher {\it et al.\/}, Phys.\ Rev.\ B {\bf 40}, 546 (1989).

\bibitem{zimanyi94}
For a brief review of the numerical work and the physical systems, see
G.T.~Zimanyi, in {\it Strongly Correlated Electronic Materials: the
Los Alamos Symposium 1933\/}, edited by K.S.~Bedell, Z.\ Huang, D.\
Smeltzer, A.\ Balatsky, and E.\ Abrahams (Addision-Wesley, New York,
1994), to appear.

\bibitem{sorensen92}
E.S.~S\o rensen, {\it et al.\/}, Phys.\ Rev.\ Lett.\ {\bf 69}, 828
(1992).

\bibitem{wallin94}
M.~Wallin, {\it et al.\/}, Phys.\ Rev.\ B {\bf 49}, 12115 (1994).

\bibitem{makivic93}
M.~Makivic, {\it et al.\/}, Phys.\ Rev.\ Lett.\ {\bf 71}, 2307 (1993).

\bibitem{weichman94b}
P.~Weichman, unpublished.

\bibitem{runge92}
K.J.~Runge, Phys.\ Rev.\ B {\bf 45}, 13136 (1992).

\bibitem{weichman94a}
Y.-H.~Tu and P.~Weichman, Phys.\ Rev. Lett.\ {\bf 73}, 6 (1994).

\bibitem{ceperley79}M.H.~Kalos, Phys.\ Rev.\ {\bf 128}, 1791 (1962);
D.M.~Ceperley and M.H.~Kalos, in {\it Monte Carlo Methods in
Statistical Physics\/}, edited by K.~Binder (Springer-Verlag,
Heidelberg, 1979), chap.~4.

\bibitem{carlson89}J.~Carlson, Phys.\ Rev.\ B {\bf 40}, 846 (1989);
N.~Trivedi and D.M.~Ceperley, Phys.\ Rev.\ B {\bf 41}, 4552 (1990);
S.~Zhang and K.J.~Runge, Phys.\ Rev.\ B {\bf 45}, 1052 (1992).

\bibitem{pollack87}
E.L.~Pollack and D.M.~Ceperley, Phys.\ Rev.\ B {\bf 36}, 8343 (1987).

\bibitem{matsubara57}
T.~Matsubara and H.~Matsuda, Prog.\ Theor.\ Phys.\ {\bf 16}, 416
(1956).

\bibitem{kawashima94}
N.~Kawashima, {\it et al.\/}, Phys.~Rev.~B, to appear; N.~Kawashima
and J.E.~Gubernatis, Phys.\ Rev.\ Lett., to appear.

\bibitem{evertz93}
H.G.~Evertz, {\it et al.\/}, Phys.\ Rev.\ Lett.\ {\bf 70}, 875 (1993).

\bibitem{wiese94}
U.-J.~Wiese and H.-P.~Ying, unpublished.

\end{references}
\end{document}